# Gold Nanoparticles Aggregation on Graphene Using Reactive Force Field: A Molecular Dynamic Study

J. Hingies Monisha[1], V. Vasumathi[1*] and Prabal K Maiti[2*],

[1]PG & Research Department of Physics, Holy Cross College (Autonomous), Affiliated to Bharathidasan University, Tiruchirappalii-620002, Tamilnadu, India.

2 Center for Condensed Matter Theory, Department of Physics, Indian Institute of Science, Bangalore-560012, India.

*Corresponding Authors: vasumathi@hcctrichy.ac.in and maiti@iisc.ac.in

**Abstract:** We examine the aggregation behavior of AuNPs of different sizes on graphene as function of temperature using molecular dynamic simulations with Reax Force Field (ReaxFF). In addition, the consequences of such aggregation on the morphology of AuNPs and the charge transfer behavior of AuNP-Graphene hybrid structure are analyzed. The aggregation of AuNPs on graphene is confirmed from the center of mass distance calculation. The simulation results indicate that the size of AuNPs and temperature significantly affect the aggregation behavior of AuNPs on graphene. The strain calculation showed that shape of AuNPs changes due to the aggregation and the smaller size AuNPs on graphene exhibit more shape changes than larger AuNPs at all the temperatures studies in this work. The charge transfer calculation reveals that, the magnitude of charge transfer is higher for larger AuNPs-graphene composite when compared with smaller AuNPs-graphene composite. The charge transfer trend and the trends seen in the number of Au atoms directly in touch with graphene are identical. Hence, our results conclude that, quantity of Au atoms directly in contact with graphene during aggregation is primarily facilitates charge transfer between AuNPs and graphene. Our results on the size dependent strain and charge transfer characteristics of AuNPs will aid in the development of AuNPs-graphene composites for sensor applications.

## 1. Introduction

The gold nanoparticles (AuNPs) and graphene family-based hybrid structure have attracted a great deal of interest because it provides additional advantages due to their improved stability and synergistic properties [1-6]. These coupled properties could lead to various potential applications such as sensors, photocatalysis, fuel cells, energy storage devices, photovoltaic devices and surface enhanced Raman Scattering [5-15].

In general, the electrical and optical properties of AuNPs depend on their size and shape [9, 16-18]. For example, there is evidence that the optical energy gap increases with decreasing AuNPs size due to the quantum confinement effect [19]. Also, recent experimental studies demonstrated that small AuNPs clusters evince more catalytic activity than larger AuNPs due to the smaller size and narrow size distribution [20 – 23]. It indicates, size is one of the crucial parameters that determines catalytic Activity of AuNPs [24-26]. In this context, some experimental studies investigated the effect of various factors such as graphene quality [27], temperature and the capping ligand [28], on morphological changes of AuNPs in AuNP-Graphene family composite. Especially, the temperature related shape changes of AuNPs are much needed for catalytic application, because the activation of metal nanoparticle catalyst requires the heat treatments such as annealing and calcination [29-31]. Such heat treatment alters the shape of AuNPs [32, 33]. For instance, Hanqing et al. [33] studied reshaping and coarsening of absorbed gold nanoparticles and nanorods on graphene oxides (GO) and their temperature dependencies. The obtained results indicate that the core size of AuNPs expands with elevated temperature. Moreover, it is reported that GO deter the stability of AuNPs by stripping the protecting ligands from the surface of AuNPs.

Since, the enhancement of sensor, catalytic and surface-enhanced Raman Spectroscopy (SERS) activities depend on the effective charge transfer between metal nanoparticles and graphene [15, 34-39]. Thus several experimental studies have focused the charge transfer behaviour of AuNP-graphene composite [15, 39- 41]. For instance, Torabi et al. [41] used the AuNPs_reduced Graphene composite as a charge transfer layer for bio-photovoltaic cells, and they observed increased overall conductivity of bio-photovoltaic cells due to the effective charge transfer between AuNPs and graphene. Very recently, Linh et al. [15] found that the inclusion of AuNP-reduced graphene oxide composite on the polydopamine (PDA) layer can improve the electron transfer capacity of PDA film, which makes this interface as an effective electrode for cytosensor to detect lung cancer cells.

The AuNPs without surface stabilizing ligands (bare_AuNPs), undergoes the aggregation to attain the stable configuration [42, 43].Further, Dutta et al. [35] studied the temperature dependant aggregation behaviour of partially bare AuNPs, and found that the rate of AuNPs aggregation increases with increasing temperature. .

Even though there have been many experimental investigations [27, 28, 33] into the aggregation of AuNPs on graphene, computational investigations into the effects of temperature and AuNP size on their structural morphology and interfacial charge transfer are lacking. Furthermore, the size of AuNPs and temperature exert the prominent effect on the catalytic activity of AuNP-Graphene composite [25, 26, 28, 33] . Hence the computational study of consequences of AuNPs aggregation on morphology of AuNPs and interfacial charge transfer of AuNP-Graphene composite are required for the development of AuNP-Graphene composite in catalytic and sensor applications. To the best of our knowledge, no simulation studies have been done in the past to address the aforementioned features.

Thus, with this motivation, here we present the MD simulation study of size and temperature dependent aggregation behaviour of AuNPs on graphene using Reactive Force Field (ReaxFF) [44]. The remaining of this paper is organized as follows: we present modelling of molecular systems and MD simulation details in next section. In section 3, we report and discuss the results from our MD simulations. In section 4, the major conclusions of this study are summarized.

## 2. Molecular Modelling and Simulation details

### 2.1. Molecular Modelling

To investigate the size dependent aggregation behavior of AuNPs on graphene, three different sizes (diameter) of AuNPs: 1.2 nm, 1.6 nm and 2.8 nm are considered for the present study. Initial configurations of these three different sized AuNPs are built using VESTA package.

The graphene sheet of size 11 × 11 nm is built using Visual Molecular Dynamics (VMD) software package. Using custom built VMD tcl script, AuNPs are arranged orderly on graphene in n x n array by maintaining the same distance between the AuNPs. Due to different sizes of AuNPs, the surface to surface distance (D) between AuNPs is slightly different (6.2, 6.3 and 6.5 Å) while maintaining the same distance between the periodic images. Further, to maintain the uniform coverage, each system consists of different n x n arrays which leads to difference in total number of AuNPs (N) arranged on graphene. The size of AuNPs, D, n x n array and the corresponding N of three systems are listed in the table 1. The initial configuration of all the systems is shown in Fig. 1.

**Table 1**

Details of various graphene-AuNPs hybrid systems simulated in this work.

| Graphene (nm$^2$) | Diameter of AuNPs (nm) | Distance between each AuNP pairs (Å) | Array (n X n) | Number of AuNPs on graphene | System Abbreviation |
|---|---|---|---|---|---|
| 11 x 11 | 1.2 | 6.2 | 6×6 | 36 | GpAu-1.2 |
| 11 x 11 | 1.6 | 6.3 | 5×5 | 25 | GpAu-1.6 |
| 11 x 11 | 2.8 | 6.5 | 4×4 | 16 | GpAu-2.8 |

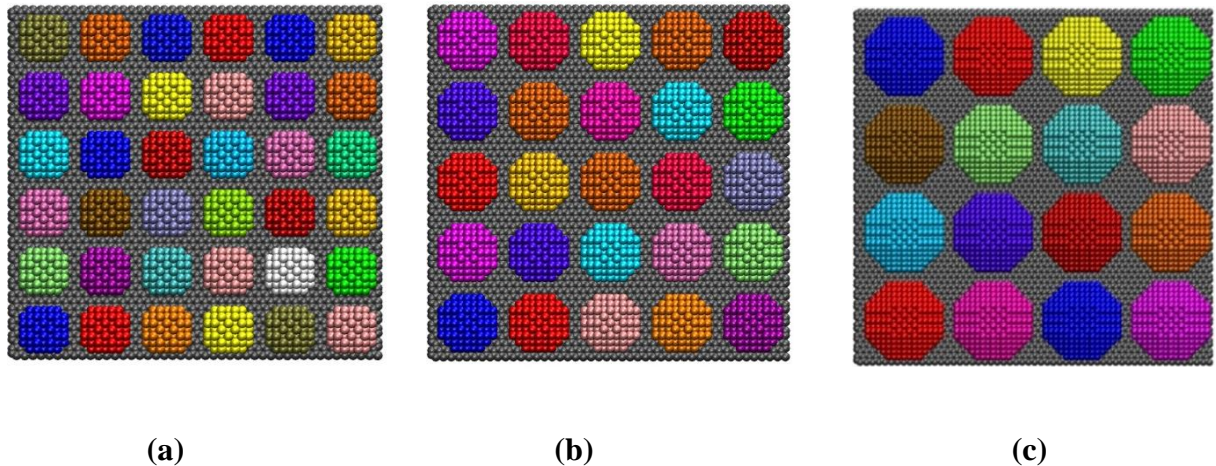

            **(a)**                              **(b)**                            **(c)**

**Fig. 1.** Snapshots of initial configuration of (a) GpAu-1.2 , (b) GpAu-1.6, and (c) GpAu-2.8. Colour coding: gray color-graphene, all other colored- different AuNPs.

**2.2 Simulation Details**

All atom MD simulation were carried out with LAMMPS software [45] by employing Reactive force field (ReaxFF) [44]. Periodic boundary conditions were applied in X and Y directions and non-periodic condition was applied in Z direction. Initially, the systems are energy minimized by using conjugate gradient method. After the energy minimization, MD simulations are performed for 1ns in the canonical ensemble (NVT) using an integration time step of 0.1fs. To investigate the effect of temperature on AuNPs aggregation, all the three systems are simulated at 300K, 400K and 500K. During simulation, temperature is kept constant by using Nose-Hoover thermostat with a relaxation constant of 1fs. All the analysis are carried out by averaging over the last 100 ps of the simulation time. The visualization of MD-trajectories is performed by VMD software package.

**2.3. Strain.** The strain of the AuNPs can be defined as,

$$\text{Strain} = \Delta L / L_o$$

Where $\Delta L = L - L_o$, $L_o$ is the original size of AuNPs and L is the size of AuNPs after aggregation.

## 3. Results and Discussion

The experimental studies reported that the performance of surface enhanced RAMAN scattering (SERS) can be improved by densely packed smaller size metal nanoparticles on graphene [27]. To probe the arrangement of AuNPs on graphene, the instantaneous snapshots of equilibrated configuration of all the three systems at different temperatures are taken and the corresponding structures are shown in Fig. 2. As seen in the Fig. 2, the AuNPs undergo aggregation for all the three cases and the pattern of aggregation varies as a function of the size of AuNPs and temperature. The GpAu-2.8 system exhibits small gap between AuNPs in aggregated clusters at all the temperatures. Whereas extremely narrow and almost no gaps between aggregated clusters are observed for GpAu-1.6 and GpAu-1.2 systems at all the temperatures. This suggests that the smaller AuNPs form tightly packed clusters compared to the larger AuNPs. As smaller size AuNPs (<1.5 nm) are less stable than larger AuNPs (~3nm) [28], it exhibits tight aggregation than that of the larger AuNPs. Fig. S. 1. shows the density contour map of AuNPs for all the systems at various temperatures. The zero contour line area in Fig. S. 1 defines the gaps between the AuNPs in the aggregated cluster. For GpAu-1.2 and GpAu-1.6 systems, minuscule gaps between the aggregated clusters are observed at all temperatures. While GPAu-2.8 system exhibits salient gaps in the aggregated cluster. It confirms the previous observation that smaller AuNPs are packed more tightly than larger AuNPs in the aggregated cluster.

Snapshots of the periodic view of each system are taken to identify the full arrangement or pattern of the aggregated cluster, which is given in Fig. S. 2. Each observed cluster pattern is unique; however, it is classified as follows:

1. Continuous pattern: The AuNPs are linked continuously (in both x and y directions) on the graphene, forming a single large cluster.
2. Partially discrete pattern: The continuous linking of AuNPs breaks (either in the x or y directions), which results in the formation of more than one cluster.
3. Long strip pattern: It is a type of partially discrete pattern where the cluster is linked like elongated strips and exhibits well-defined gaps between them.

The GpAu-1.2 system exhibits a continuous aggregated pattern on the basal plane of graphene at 300K. Besides, the GpAu-1.6 and GpAu-2.8 systems showed a prominent long strip aggregated pattern. Especially, the aggregated pattern of GpAu-2.8 is well ordered compared with the GpAu-1.6 system. It is clearly observed that as the size of the AuNPs increased, the pattern of the aggregated cluster became well ordered. Furthermore, it is also noticed that all the systems show changes in pattern at higher temperatures. For example, the GpAu-1.2 system at 400K exhibits a partially discrete aggregated pattern, whereas at 500K it shows a continuous aggregated pattern. As the temperature increased to 400 K, a continuous aggregation pattern was observed for the GPAu-1.6 and GpAu-2.8 systems. Furthermore, at 500K, the GPAu-1.6 system shows partially discrete pattern and GPAu-2.8 system exhibit a long strip pattern along the diagonal axis. The ordered pattern of aggregated clusters is mostly observed for larger AuNPs at lower temperatures. While, for higher temperatures, the continuous and partially discrete pattern is more optimal. For very small AuNPs, a continuous aggregated pattern is mostly observed at all the temperatures.

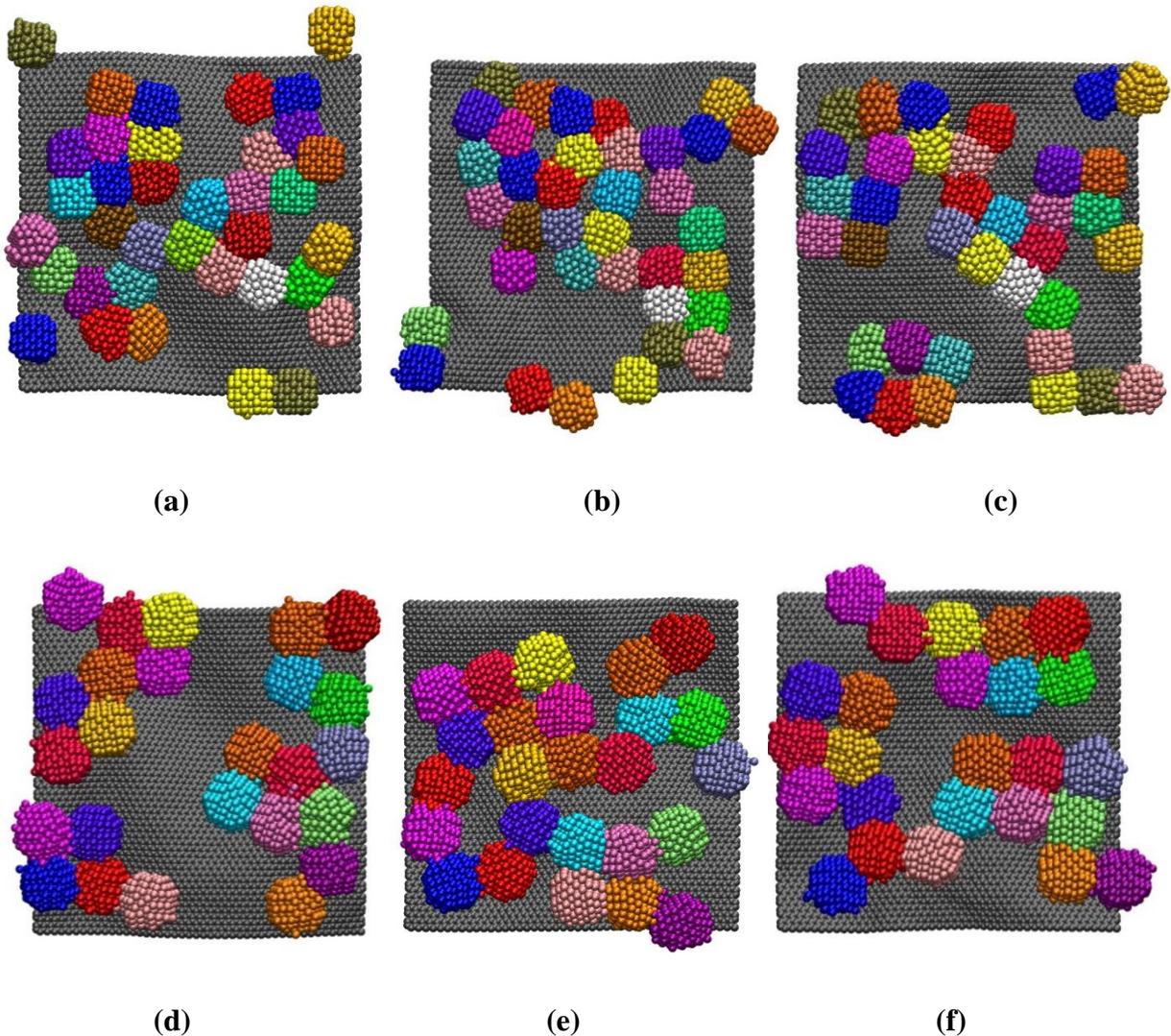

(a)  (b)  (c)

(d)  (e)  (f)

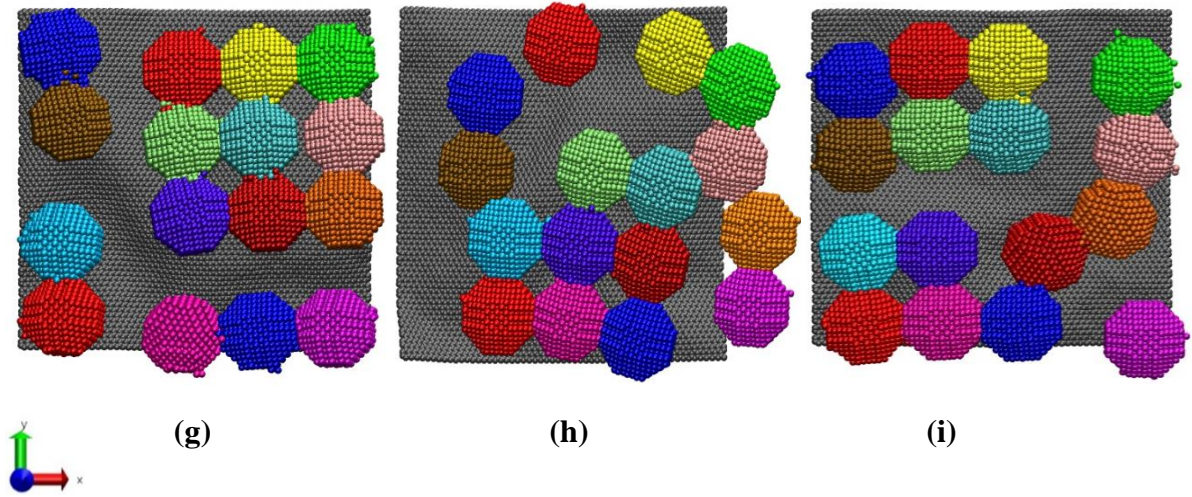

**(g)**                  **(h)**                  **(i)**

**Fig. 2.** Snapshots of final configuration of GpAu-1.2 at (a) 300K, (b) 400K, (c) 500K, GpAu-1.6 at (d) 300K, (e) 400K, (f) 500K, GpAu-2.8 at (g) 300K, (h) 400K and, (i) 500K.

### 3.1 Center of Mass Distance

To understand the aggregation behaviour of AuNPs on graphene, the center of mass distance (d) between each pair of AuNP is calculated (see Fig. 3). All nearest pairs are considered for the center of mass distance calculation. The initial center of mass distance ($d_{initial}$) between AuNP pairs for GpAu-1.2, GpAu-1.6 and GpAu-2.8 systems are 18, 22 and 34 Å, respectively. Since, the distance (D) between AuNP pairs is kept constant, $d_{initial}$ value is same for within the system. The d value between AuNP pairs is calculated and the corresponding probability distribution for all the cases are shown in Fig. 4. To find the pairs that involve in aggregation, we have calculated closest distance between AuNP pairs ($d_c$) value by employing the below formula

$$d_c = r_1 + 2.88\text{Å} + r_2$$

Where, 2.88 Å is the optimised distance between Au Atoms [46]. $r_1$ and $r_2$ are radius of 1st and 2nd AuNP in a pair, repectively.

The calculated $d_c$ values are 14.88, 18.88 and 30.88 Å for GpAu-1.2, GpAu-1.6 and GpAu-2.8 system, respectively. This $d_c$ value serves as the threshold to determine if a pair will participate in aggregation or not. In all the systems, the distribution observed at d ≤ $d_c$ and d > $d_c$ correponds to aggregation and dispersion of AuNPs on graphene (see Fig. 3). The observed maximum distribution at d ≤ $d_c$ for all the systems, clearly represents many AuNPs are involved in the aggregation. The minimum distribution at d > $d_c$ represents the dispersion of few AuNPs as well as formation of different group of clusers.

As seen in Fig. 4, the probability distribution of AuNPs is almost continuous for the GpAu-1.2 system. For the GpAu-1.6 system, the distribution is slightly discrete. whereas a more discontinuous distribution is observed for the GpAu-2.8 systems. These diverse distributions suggest that there are distinct ways for AuNPs to aggregate on graphene sheet. In other words, for GPAu-1.2 system, almost all the AuNPs involve tight aggregation and exhibit almost continuous distrbution. On the otherhand, for GpAu-1.6 and GPAu-2.8 systems (see Fig. 2), group of AuNPs clusters are observed that leads the discrete distribution. Comparing to other two systems, the GpAu-2.8nm system exhibits most ordered aggregation, which results in the largest distribution at $d \leq d_c$. For all the systems, when temperature rises, the maximum distribution at $d \leq d_c$ get lowers. These size and temperature dependent variation in distribution pattern, clearly indicates the formation of different pattern of aggregated cluster on graphene. This observation is in line with a conclusion made from the previous section. Next, descriminate the disctribution at $d = d_c$ and $d < d_c$.

The distribution at $d < d_c$ and $d = d_c$ corresponds to shape changes and no shape changes of AuNPs during the aggregation, respectively. As seen in Fig. 4, in all the systems, the highest peak observed at $d = d_c$ as well as later highest peaks noticed at $d < d_{close}$. It depicts, there was a change in shape of AuNPs during the aggregation.

The range of disribution at $d < d_c$ measures the amount of AuNPs compression during the aggregation (see Fig. 3). A wider distribution at $d < d_c$ implies the higher degree of AuNPs compression and the narrow spread suggests the lower degree of compression. Thus, we can state that the degree of AuNPs compression is increased with a size decrement of AuNPs. In other words, the smaller size AuNPs exhibit more compression compared to the case of larger AuNPs during aggregation, as smaller size AuNPs are more tightly aggregated compared with larger AuNPs. As can be seen from Fig. 4, at all tempertures, GpAu-1.2 and GpAu-1.6 systems exhibit wider distribution when comapared with GpAu-2.8 system. Also, we observed that the distribution of GPAu-1.2 system is slightly wider compared to GpAu-1.6 at 400K and 500K temperature. The range of spread at $d < d_c$ is almost equal for GpAu-2.8 systems at different temperatures. Whereas, GpAu-1.6 system exhibits slightly wider distribution at 300K when compared with higher tempertures and the range of spread is almost equal for 400K and 500K. The range of spread at $d < d_c$ is slightly wider for GpAu-1.2 system at 300K and 500K compared to 400K. This suggest that the degree of compression of AuNPs slightly varies with temperature but does not increase linearly with temperature.

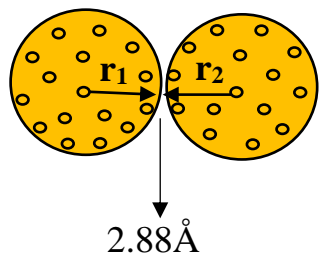

(a)

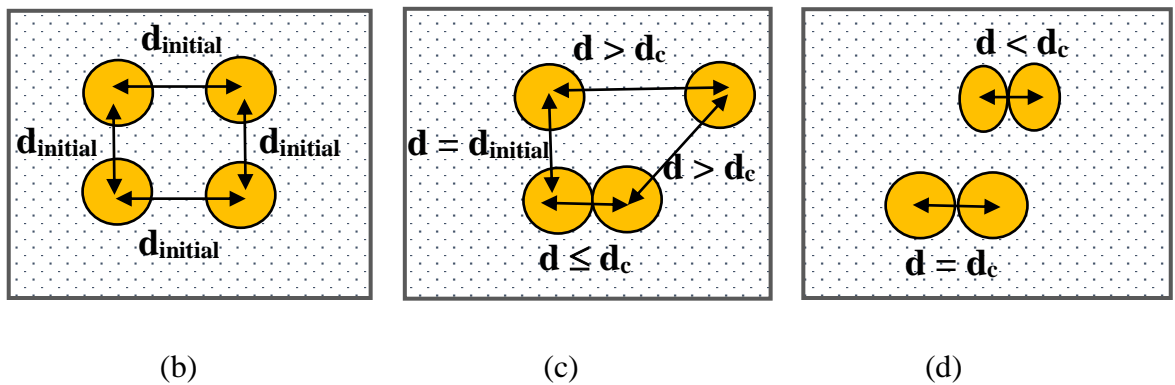

(b)    (c)    (d)

**Fig. 3**. Two-dimentional schematic representaion of the (a) closest center of mass distance calculation between AuNPs, (b) initial center of mass, (c) aggregation and dispersion, and (d) compression of AuNPs. large/medium yellow circle and dotted square refer to AuNP and grapehene respectively. Small yellow circle refers to Au atom.

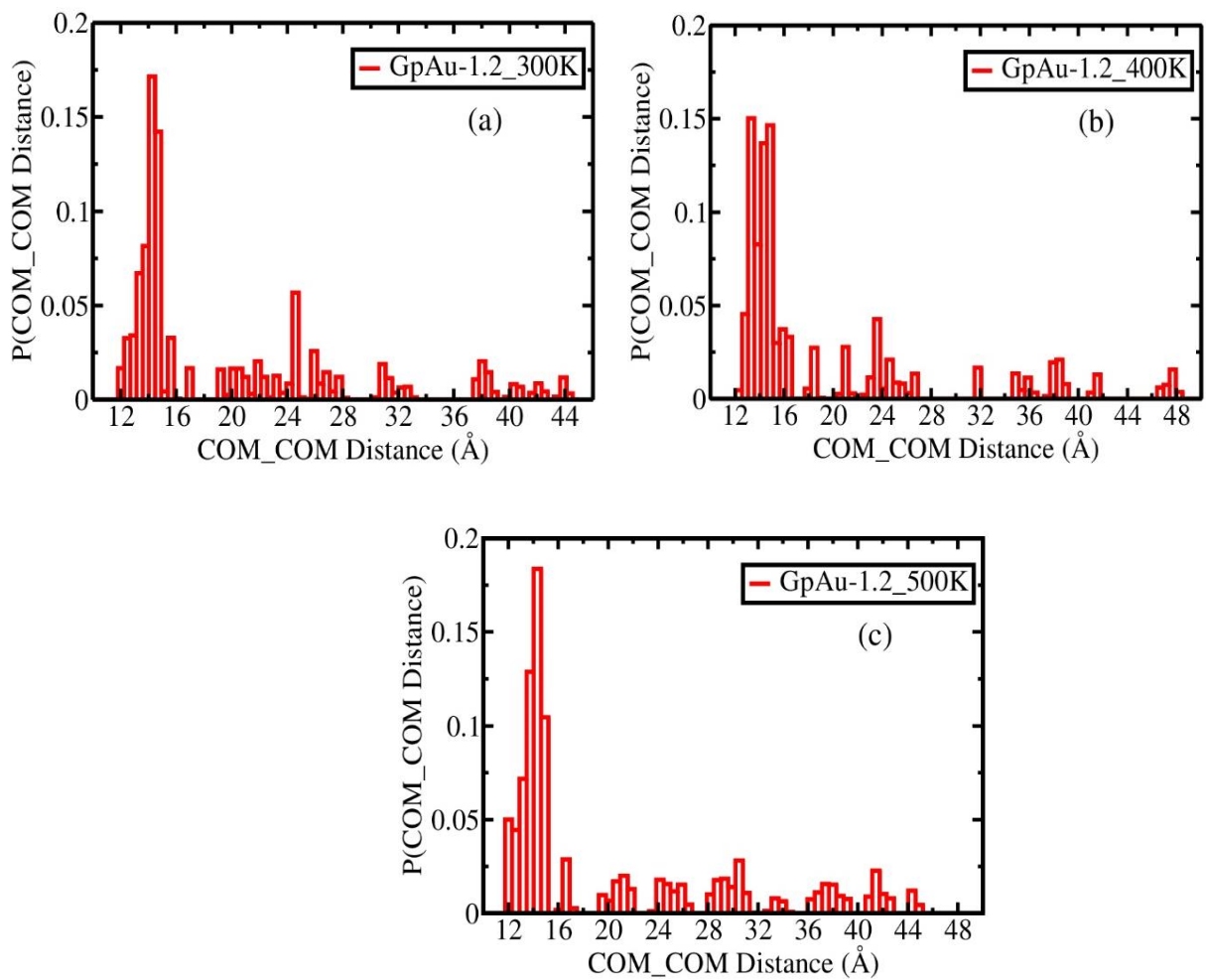

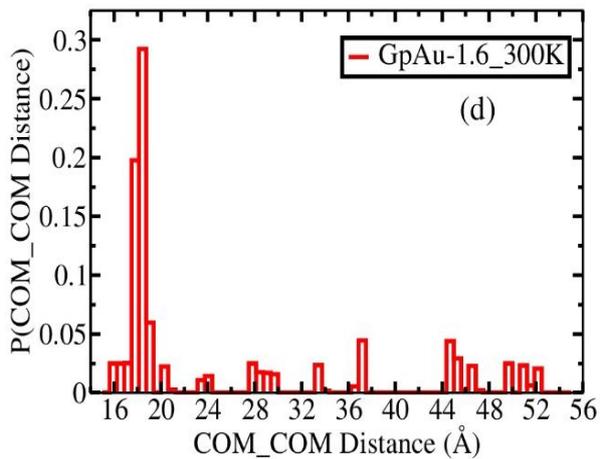
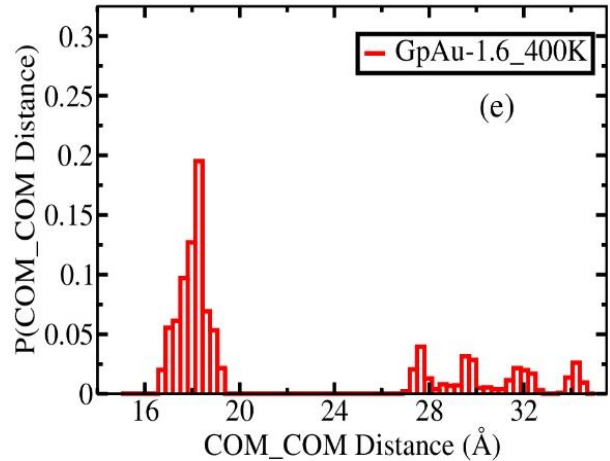
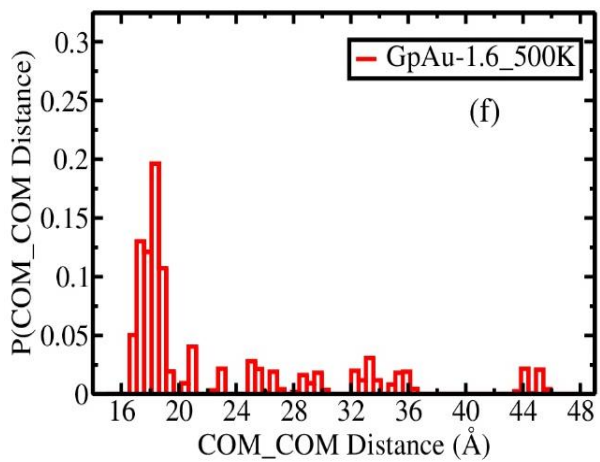
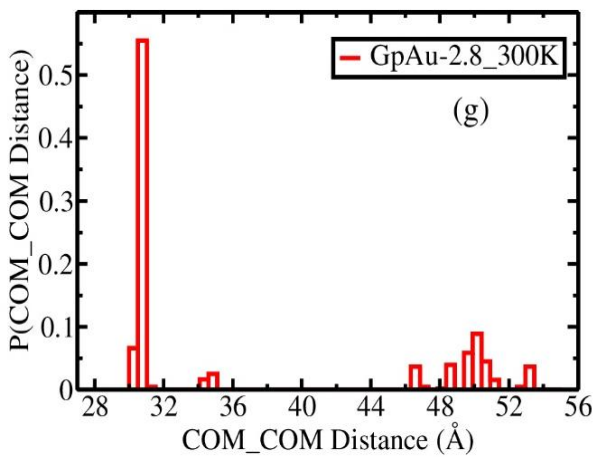
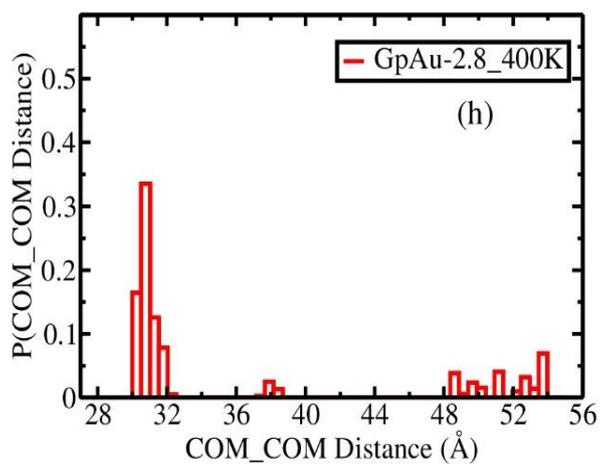
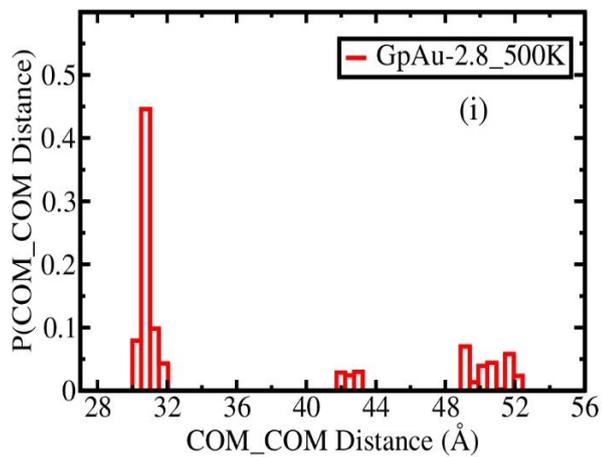

**Fig. 4.** The center of mass distance between AuNPs pairs GpAu-1.2 system at (a) 300K, (b) 400K, (c) 500K , GpAu-1.6 system at (d) 300K, (e) 400K, (f) 500K, GpAu-2.8 system at (g) 300K, (h) 400K and (i) 500K.

### 3.2 Strain of AuNPs

The catalytic activity and the surface plasmon resonance of AuNPs is highly influenced by their structural morphology (size and shape) [47 - 49] The COM distribution profile of AuNPs exhibits some of the AuNPs are compressed during aggregation. To probe this observation as well as to find the other shape changes occurs along with the compression, we have calculated the strain of AuNPs along principal axis for all the systems. To recognize easily, the identification number (ID) for each AuNPs are assigned and it is allocated in a manner that, starting from 1 in bottom left to right in a first row then right to left in second row and repeated up to the corresponding N value which is illustrated in Fig. 5. The strain of each AuNPs in all three directions with a corresponding identification number is shown in Fig. 6. The positive and negative strain values represent the enlargement and compression of AuNPs, respectively. The significant variations in strain observed in all the systems depict that each AuNPs undergoes enlargement and compression in all the three directions. The enlargement and compression of AuNPs arises from the collision of AuNPs during the aggregation. In other words, the strain of each AuNP depends on where it is located in the aggregated cluster and how tightly it is aggregated with the other AuNPs in the system.

As seen in Fig. 6, AuNPs of GpAu-1.2 exhibits a higher strain than the other two systems. Further, in GpAu-2.8 system, the AuNPs evince lower magnitude of enlargement and almost no compression. As AuNPs of GpAu-1.2 system tightly aggregated compare with GpAu-1.6 system, it shows larger enlargement and compression than that of the GpAu-1.6 system. However, AuNPs of GpAu-2.8 system is not tightly aggregated as compared with the smaller AuNPs systems, which results in relatively very low enlargement and compression. Also, each system shows a non-identical strain profile at different temperature. For instance, several AuNPs in GpAu-1.2 system at 500K show a greater degree of compression and enlargement compared with the other two temperatures. Whereas, the higher degree of compression and enlargement of AuNPs observed at 300K for GpAu-1.6 and GpAu-2.8 systems. This size and temperature dependent variation in strain is associated with a pattern of the aggregated cluster.

The variation in strain of the individual AuNPs within the system is also related to the pattern of the aggregated cluster; that is, the AuNPs in a tightly aggregated part of the cluster yield more strain or shape changes, when compared to the other parts of aggregated cluster. This can be verified by comparing the strain profile with a snapshots of the equilibrated configuration of all the systems. For example, in case of GpAu-1.2 at 300K, the AuNP of ID-27 exhibits highest enlargement and compression as it is located at the tightly aggregated part of the cluster. Whereas ID-5 shows almost no compression and enlargement, because it is not positioned in the tightly aggregated part of cluster (see Fig. S. 3). The same trend can be seen in all other systems.

To elaborate clearly about the shape changes of AuNPs, the snapshots of the highly enlarged and compressed AuNP of each system with a corresponding ID number is presented in Fig. 6. The AuNPs of GpAu-1.2 and GpAu-1.6 systems shows the shape changes at all the temperatures, while in GpAu-2.8 system, only dislocation of Au atoms is observed. Previous experimental study on the morphological transformation of functionalised AuNPs on graphene oxide reported that smaller AuNPs undergoes slightly more coarsening compare with larger AuNPs (~3nm) on graphene oxide [28]. This qualitative observation also illustrates the same, that is, smaller AuNPs underwent more shape modification during the aggregation than the larger AuNPs.

To quantify the degree of AuNPs compression and enlargement, we have calculated the average strain, that is, average enlargement and compression of all the AuNPs in all the systems and the results are presented in Fig. 7. From the Fig. 7, we infer that the average compression and enlargement of AuNPs increased with a size decrement of AuNPs in all the cases. For GpAu-1.2 system, the average compression and enlargement of AuNPs are increased significantly with a temperature increment. While, GpAu-1.6 system at 500K and 300K exhibit slightly higher average compression and enlargement than at 400K. In contrast, in the GpAu-2.8 system, an increase of temperature results in a very modest decrease in the average compression and enlargement of AuNPs.

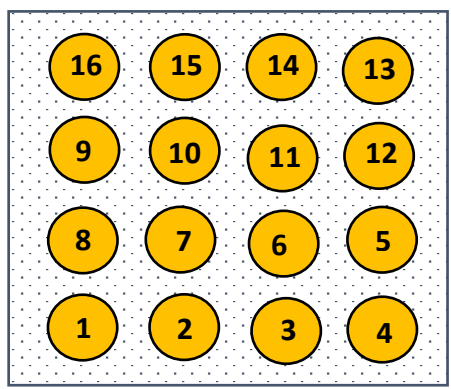

**Fig. 5.** A two-dimensional schematic representation of AuNPs on graphene with a corresponding identification number.

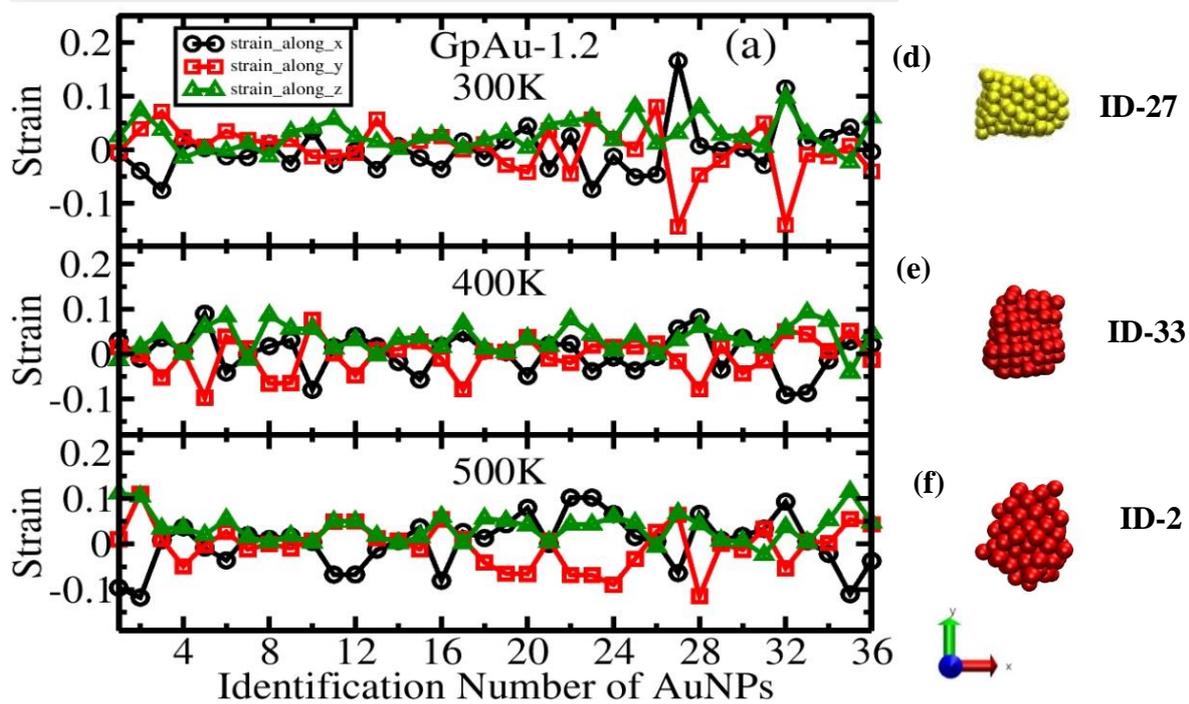
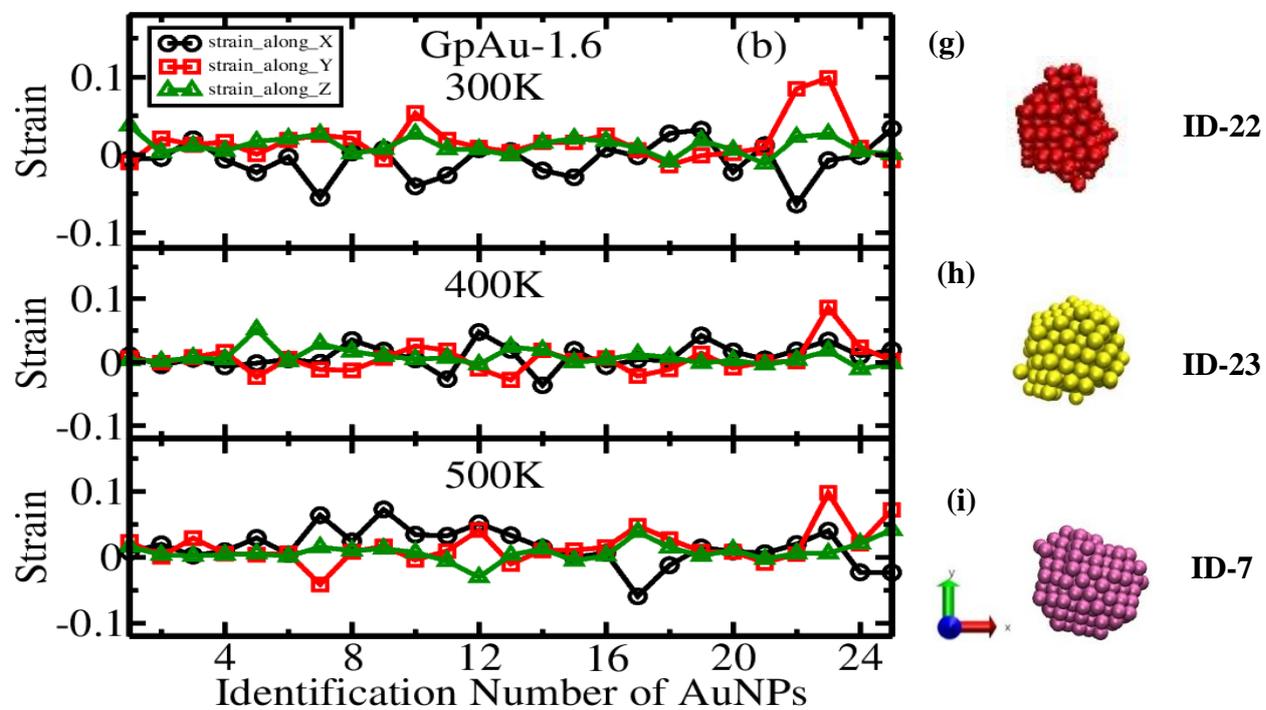

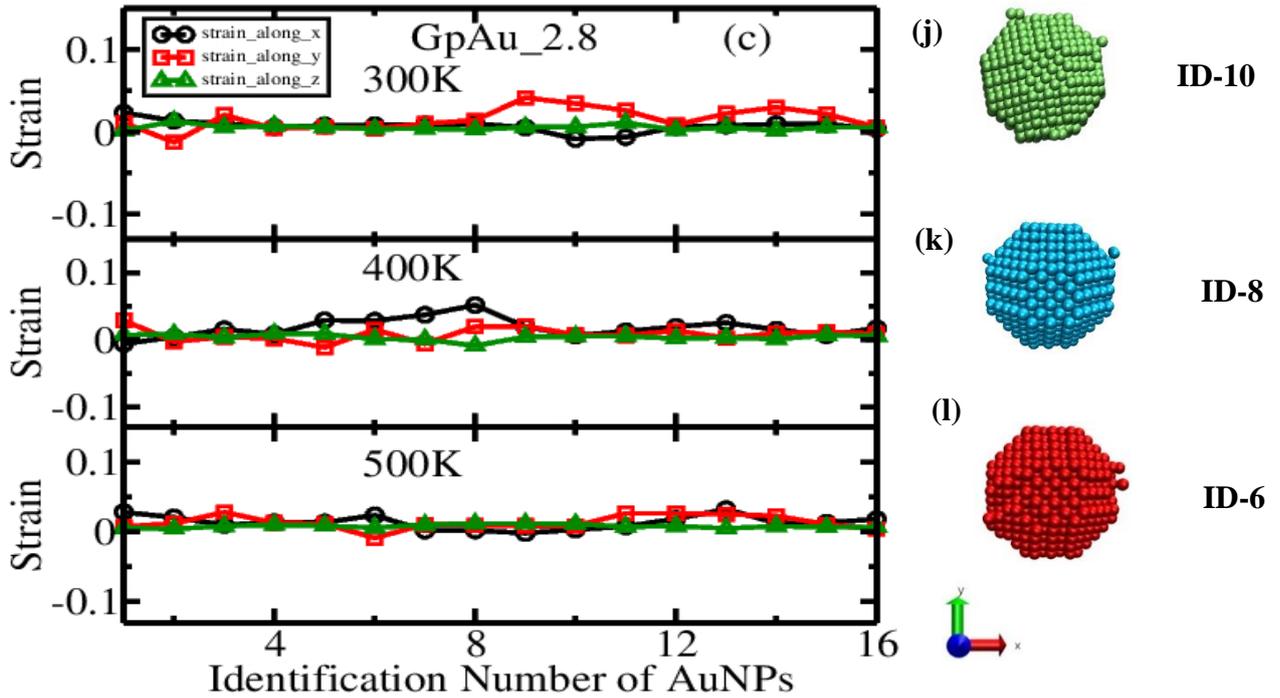

**Fig. 6.** The average strain of each AuNPs in X, Y and Z direction for last 100ps of the simulation. The left column of figures corresponding to (a) GpAu-1.2, (b) GpAu-1.6, (c) GpAu-2.8 system at 300, 400 and 500K. Right column corresponds to the highly shape modified AuNPs of GpAu-1.2 at (d) 300K, (e) 400K, (f) 500K, GpAu-1.6 at (g) 300K, (h) 400K, (i) 500K, GpAu-2.8 at (j) 300K , (k) 400K and (l) 500K .

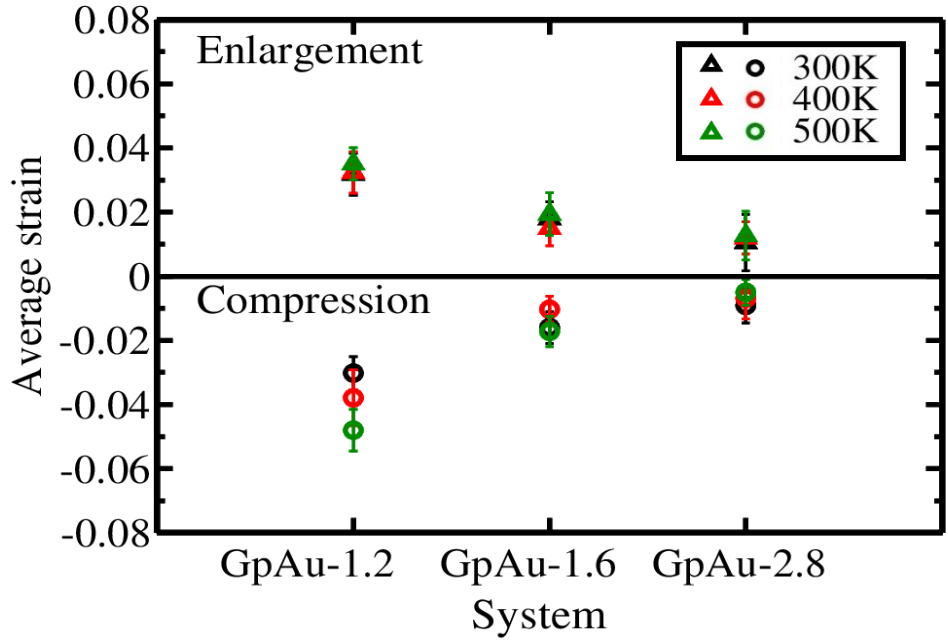

**Fig. 7.** The average strain of all the AuNPs in each system.

### 3.3 Charge Transfer

One of the significant descriptors for the catalytic activity of metal nanoparticle-graphene composite is effective electron transport from the metal nanoparticle to graphene [50]. The experimental studies on the AuNP-Graphene composite, observed the occurrence of charge transfer between the AuNPs and the graphene [7, 51-53]. To understand the catalytic activity, the charge transfer between AuNPs and graphene has been studied for all the cases. Simulation using ReaxFF allows to recalculate the charge of each and every atom by QeQ method [54]. The total charge of AuNPs and graphene as a function of simulation time is shown in Fig. S. 3 of the supporting information. The variation in total charge during the simulation time corresponds to the charge transfer between AuNPs and graphene. In other words, increasing positive and negative charge of AuNPs and graphene, respectively, depicts the transfer of electron from AuNPs to graphene. As seen in Fig. S. 4, in all the systems, the AuNPs and graphene show increased positive and negative total charge, which clearly illustrate the occurrence of charge transfer between AuNPs and graphene.

To quantify the interfacial charge transfer of all the cases, we have calculated the amount of charge transfer by using below formula

$$\text{Charge transfer} = Q_f - Q_i$$

Where, $Q_f$ and $Q_i$ are the final and initial total chare of AuNPs/graphene repectively.

The modulus of amount of charge transfer for all the systems is summarized in Table 2. It is found that the amount of the electron transfer of metal nanoparticle-graphene composite depends on the size of nanoparticles [49]. From the Table 2, it is observed that for some systems the amount of charge transfer increased with size increment of AuNPs. For example, at all the temperatures, the GpAu-2.8 system, exhibits higher amount of the charge transfer than other two systems. While, GpAu-1.6 system exhibits slightly higher charge transfer than GpAu-1.2 system only at 400K temperature, for remaining temperatures it is vice versa. This contrary can be explained by the smaller difference between the AuNPs size of both GpAu-1.2 and AuNP-1.6nm systems. Also the amount of charge transfer is slightly increased with a temperature increment, in GpAu-1.2 and GpAu-2.8 systems. However, in GpAu-1.6 system, the highest charge transfer is observed at 400K temperature. This inconsistency might be associated with a variation in the aggregated pattern of AuNPs on graphene.

A visual inspection of MD trajectory reveals that during the aggregation some AuNPs levitated from graphene. The levitation of AuNPs arises from the aggregation, which is related to the pattern of aggregated cluster. The AuNPs that are completely and partially levitated from graphene are shown in Fig. S.5. This levitation of AuNPs can minimize the direct contact of AuNPs with graphene, which may affect the charge transfer between AuNP and graphene through conduction.

To examine the correlation between aggregated cluster pattern and interfacial charge transfer of AuNP-Graphene composite, we have calculated the number of Au atoms directly in contact ($N_c$) with a graphene (Au atoms within 4 Å from Graphene) by using our own VMD tcl script. Because of the varied size and number of AuNPs, the number of Au atoms directly in contact with the graphene for different sized systems will also be different. Table 3, portrays the $N_c$ of all the systems, where it is observed that the $N_c$ values in each system slightly vary with temperature. Table 3, illustrates that in GpAu-1.2 and GpAu-1.6 systems $N_c$ values grows linearly as temperature rises. Whereas, GpAu-1.6 system at 400K exhibit higher $N_c$ value compare with other two temperatures. By comparing this observation with that the obtained values from Table 2, it is revealed that the amount of charge transfer is directly associated with $N_c$ values.

To summarize, our results indicate that the amount of charge transfer between AuNPs and graphene in AuNP-Graphene composite depends on size of AuNPs, temperature and the number of Au atoms directly in contact with graphene. Further, it is found that increasing the size of AuNPs facilitates the charge transfer and thus the catalytic activity can be enhanced.

**Table 2**

**The average magnitude of charge Transfer between AuNPs and graphene.**

| System name | Modulus of charge transfer (C) | | |
|---|---|---|---|
| | **300K** | **400K** | **500K** |
| GpAu-1.2 | 14.34 ± 0.18 | 14.94 ± 0.22 | 14.96 ± 0.17 |
| GpAu-1.6 | 13.79 ± 0.14 | 15.14 ± 0.20 | 14.12 ± 0.25 |
| GpAu-2.8 | 20.73 ± 0.21 | 20.78 ± 0.23 | 21.71 ± 0.18 |

**Table 3**

**The average number of Au atoms directly in contact with graphene**

| System name | Number of Au atoms directly in contact with graphene | | |
|---|---|---|---|
| | **300K** | **400K** | **500K** |
| GpAu-1.2 | 333.20 ± 3.94 | 347.21 ± 4.12 | 349.02 ± 4.24 |
| GpAu-1.6 | 276.17 ± 4.77 | 304.39 ± 4.53 | 283.24 ± 3.43 |
| GpAu-2.8 | 386.13 ± 4.39 | 392.23 ± 3.21 | 403.29 ± 3.09 |

**4. Conclusion**

In this study, molecular dynamic simulation is used to examined the size and temperature dependent aggregation behavior of AuNPs on graphene. We also investigated how AuNPs aggregation affect the shape of individual AuNPs and the charge transfer between AuNPs and graphene. Equilibrated simulated structures show that the temperature and AuNP size affect the pattern of AuNP aggregation on graphene. The results from center of mass distribution show that aggregates of smaller size AuNPs are very tight whereas the larger size AuNPs shows gapes in aggregated cluster in all the temperatures. Further, the distribution at $d<d_c$ depicts that AuNPs undergoes shape modifications during aggregation.

The strain calculation show that, the majority of AuNPs exhibit changes in shape due to the collision of AuNPs during the aggregation. The magnitude of strain (both enlargement and compression) is increase with decrease of size of AuNPs i.e. when compared to larger AuNPs, smaller size AuNPs display significantly more shape alterations. The reason behind more shape alterations for smaller size AuNPs is that smaller size AuNPs are tightly aggregated with other AuNPs, whereas larger size AuNPs are not tightly aggregated.

The calculation of charge transfer show that the electrons are transferred from the AuNPs to graphene. The larger size AuNPs transfers more electron to graphene compare with smaller size AuNPs at all the three temperatures. Within the same system, there exist a small variation in charge transfer at different temperature, but the charge transfer is not linearly increased or decreased with temperature. This observed variance is due to the number of Au atoms directly in contact with the graphene. In other words, some AuNPs are pushed away from graphene during aggregation, preventing them from coming into close contact with the graphene and reducing charge transfer. As a result of this, our results show that the quantity of Au atoms directly in touch with graphene is primarily responsible for facilitating the transfer of charges between AuNPs and graphene.

In conclusion, the results from center of mass distribution and strain of AuNPs suggest that the lower size AuNPs are ideal for rough/disordered surfaces whereas larger size AuNPs are required for ordered coatings. We believe that the present study provides a better understanding of the size dependent AuNPs aggregation behavior on graphene and charge transfer between AuNPs and graphene. Further, the obtained charge transfer results show that charge transfer between AuNPs and graphene exclusively depends on the arrangements of AuNPs during aggregation along with size of the AuNPs. Hence, our findings of charge transfer and aggregated pattern of AuNPs pave a way to design and engineering the AuNPs-graphene based sensors. Our results show that the structural properties and interfacial charge of the AuNPs during aggregation are more influenced by their size than by temperature. Although the current work only examines bare AuNPs, experimental studies have shown that temperature has a substantial impact on the aggregation of functionalized AuNPs [28, 33, 35]. It will be examined to see if functionalized AuNPs significantly alter the way that temperature affects aggregation. In addition, our strain calculation results suggest that smaller AuNP sizes exhibit more strain when compared to larger AuNP sizes. The intriguing question is whether this strain is decreased when we use functionalized AuNPs. Studies targeting similar problems are now being conducted in our laboratory.

**Acknowledgement**

V.V. acknowledges the financial support from SERB-TARE (TAR/2018/001354).

**References**

1. Y. Choi, H. S. Bae, E. Seo, S. Jang, K. H. Park, B. –S. Kim, Hybrid gold nanoparticle-reduced graphene oxide nanosheets as active catalysts for highly efficient reduction of nitroarenes. *J.Mater.chem.* 21, 15431-15436 (2011).

2. P. T. Yin, T. –H. Kin, J. –W. Choi, K. –B. Lee, Prospects for graphene –nanoparticle-based hybrid sensors. *Phys. Chem. Chem. Phys*. 15, 12785-12799 (2013).

3. S. Basu, S. K. Hazra, Graphene-Noble Metal Nano-Composites and Applications for Hydrogen Sensors. *C.* 3, 29 (2017).

4. V. Amendola, R. Pilot, M. Frasconi, O. M. Maragò, M. A. Iatì, Surface plasmon resonance in gold nanoparticles: a review. *J. Phys. Condens Matter.* 29 , 203002 (2017).

5. S. Sun, P. Wu, Easy Fabrication of Macroporous Gold Films Using Graphene Sheets as a Template. *ACS Appl. Mater.  Interfaces.* 5, 3481-3486 (2013).

6. H. Ismaili, D. Geng, A. X. Sun, T. T. Kantzas, M. S. Workentin, Light-Activated Covalent Formation of Gold Nanoparticles-Graphene and Gold Nanoparticle-Glass Composites. *Langmuir.* 27, 13261-13268 (2011).

7. Y. Ren, R. Rao, S. Bhusal, V. Varshney, G. Kedziora, R. Wheeler, Y. Kang, A. Roy, D. Nepal, Hierarchical Assembly of Gold Nanoparticles on Graphene Nanoplatelets by Spontaneous Reduction: Implications for Smart Composites and Biosensing. *ACS Appl. Nano Mater.* 3, 8753-8762 (2020).

8. H. Yin, H. Tang, D. Wang, Y. Goa, Z. Tang, Facile Synthesis of Surfactant-Free Au Cluster/Graphene Hybrids for High-Performance Oxygen Reduction Reaction. *ACS Nano.* 6, 8288-8297 (2012).

9. C. Tan, X. Huang, H. Zuang, Synthesis and Applications of Graphene-based Nobel Metal Nanostructures. *Mater.Today*.16, 29-36 (2013).

10. W. Hong, H. Bai, Y. Xu, Z. Yao, Z. Gu, G. Shi, Preparation of Gold Nanoparticle/Graphene Composites with Controlled Weight Contents and Their Applications in Biosensors. *J. Phy. Chem. C*. 114, 1822-1826 (2010).

11. O. –A. Lazer, A. Marinoiu, M. Raceanu, A. Pantazi, G. Mihai, M. Varlam, M. Enachesuc, Reduced Graphene oxide decorated with dispersed gold nanoparticles: Preparation, Characterisation and electrochemical evaluation for Oxygen reduction reaction. *Energies.* 13, 4307 (2020).

12. S. Rattan, S. Kumar, J. K. Goswamy, Gold nanoparticle decorated graphene for efficient sensing of $NO_2$ gas. *Sensors international* 3, 100147 (2022).

13. L. G. Baumgarten, A. A. Freitas, E. R. Santana, J. P. Winiarski, João, J. P. Dreyer, I. C. Vieira, Graphene and gold nanoparticle-based bionanocomposite for the voltammetric determination of bisphenol A in (micro)plastics. *Chemosphere.* 334, 139016 (2023).

14. D. P. Kepić, A. M. Stefanović, M. D. Budimir, V. B. Pavlović, A. Bonasera, M. Scopelliti, B. M. Todorović-Marković, Gamma rays induced synthesis of graphene oxide/gold nanoparticle composites: structural and photothermal study. *Radiation Physics and Chemistry.* 202, 110545 (2023).

15. N. D. Linh, N. T. T. Huyen, N. H. Dang, B. Piro, V. Thi Thu, Electrochemical interface based on polydopamine and gold nanoparticles/reduced graphene oxide for impedimetric detection of lung cancer cells. *RSC Adv.* 13, 10082-10089 (2023).

16. J. Zhang, L. Mou, X. Jiang, Surface chemistry of gold nanoparticles for health-related applications. *Chem. Sci.* 11, 923-936 (2020).

17. M. A. H. Khalafalla, A. Mesli, H. M. Widattallah, A. Sellai, S. H. Al-harthi, H. A. J. Al-Lawati, F. O. Suliman, Size-dependent conductivity dispersion of gold nanoparticle colloids in a microchip: contactless measurements. *J.Nanoparticle Res*. 16, 2546 (2014).

18. E. S. Kooij, W. Ahmed, H. J. W. Zandvilet, B. Poelsema, Localized Plasmons in Noble Metal Nanospheroids. *J.Phys. chem.C.* 155, 13021-10332 (2011).

19. G. M. A. Gad, M. A. Hegazy, Optoelectronic properties of gold nanoparticles synthesized by using wet chemical method. *Mater. Res. Express*. 6, 2053-1591 (2019).

20. W. Chen, S. Chen, Oxygen Electroreduction Catalyzed by Gold Nanoclusters: Strong Core Size Effects. *Angew. Chem., Int. Ed*. 48, 4386-4389 (2009).

21. C. Jeyabharathi, S. S. Kumar, G. V. Kiruthika, K. L. N. Phani, Aqueous CTAB-Assisted Electrodeposition of Gold Atomic Clusters and Their Oxygen Reduction Electrocatalytic Activity in Acid Solutions. *Angew.Chem.,Int.Ed.Engl*. 49, 2925-2928 (2010).

22. N. Kapil, F. Cardinale, T. Weissenberger, T. A. Nijhuis, M. M. Nigra, M. –O. coppens, Gold nanoparticles with tailored size through ligand modification for catalytic applications. .*Chem.Commun.* 57 10775-10778 (2021).

23. P. Suchomel, L. Kvitek, R. Prucek, A. Panacek, A. Halder, S. Vajda, R. Zboril, Simple size-controlled synthesis of Au nanoparticles and their size-dependent catalytic activity. *Sci, Rep.* 8, 4589 (2018).

24. B. G. Donoeva, D. S. Ovoshchnikov, V. B. Golovko, Establishing a Au Nanoparticle Size Effects in the Oxidation of Cyclohexene Using Gradually Changing Au Catalysts. *ACS Catal.* 3, 2986-2991 (2013).

25. N. T. Khoa, S. W. Kim, D. –H. Yoo, E. J. Kim, Size-dependent work function and catalytic performance of gold nanoparticles decorated graphene oxide sheets. *Applied Catalysis A: General*. 469, 159- 164 (2014).

26. Q. Yang, M. Dong, H. Song, L. Cao, Y. Zhang, L. Wang, P. Zhang, Z. Chen, Size dependence electrocatalytic activity of gold nanoparticles decorated reduced graphene oxide for hydrogen evolution reaction. *J Mater Sci Mater Electron.* 28, 10073–10080 (2017).

27. T. Kostadinova, N. Politakos, A. Trajcheva, J. Blazevska-Gilev, R. Tomovska, Effect of Graphene characteristics on Morphology and Performance of Composite Noble Metal-Reduced Graphene Oxide SERS substrate. *Molecules* 26, 4775 (2021).

28. H. Pan, S. Low, N. Weerasuriya, B. Wang, Y. –S. sho, Morphological transformation of gold nanoparticles on graphene oxide: effect of capping ligands and surface interaction. *Nano Covergence.* 6, 2 (2019).

29. M. Du, D. Sun, H. Yang, J. Hung, X. Jing, T. Odoom-wubha, H. Wang, L. Jia, Q. Li, Influence of Au Particle Size on Au/$TiO_2$ Catalysts for CO Oxidation. *J. Phy. Chem. C.* 118, 19150-19157 (2014).

30. L. Adijanto, A. Sambath, A. S. Yu, M. Cargnello, P. Fornasiero, R. J. Gorte, J. M. Vohs, Synthesis and Stability of Pd@$CeO_2$ Core-Shell Catalyst in solid Oxide Fuel Cell Anodes. *ACS Catal.* 3, 1801-1809 (2013).

31. M. W. Sugden, T. H. Richardson, G. Leggett, Sub-10 Ω Resistance Gold Films Prepared by Removal of Ligands from Thiol-Stabilized 6 nm Gold Nanoparticles. *Langmuir* 26, 4331-4338 (2010).

32. J. Hrbek, F. M. Hoffmann, J. B. Park, P. Liu, D. Stacchiola, Y. S. Hoo, S. Ma, A. Nambu, J. A. Rodriguez, M. G. White, Adsorbate-Driven Morphological Changes of a Gold Surface at Low Temperature. *J. Am. Chem. Soc.* 130, 17272-17273 (2008).

33. H. Pan, S. Low, N. Weerasuriya, Y. –S. Shon, Graphene Oxide-Promoted Reshaping and Coarsening of Gold Nanorods and Nanoparticles. *ACS Appl. Mater. Interfaces.* 7, 3406-3413 (2015).

34. M. Khavani, M. Izadyar, M. R. Housaindokht, Modeling of the Functionalized Gold Nanoparticle Aggregation in the presence of Dopamine: A Joint MD/QM Study. *J. Phys. Chem. C.* 112, 26130-26141 (2018).

35. A. Dutta, A. Paul, A. Chattopadhyay, The effect of temperature on the aggregation kinetics of partially bare gold nanoparticles. *RSC Adv.,* 6, 82138-82149 (2016).

36. M. Zhou, A. Zhang, Z. Dai, Y. P. Feng, C. Zhang, Strain-Enhanced Stabilization and catalytic Activity of Metal Nanoclusters on Graphene. *J. Phys. Chem. C.* 114, 16541-16546 (2010).

37. D –H. Lim, J. Wilcox, DFT-Based Study on Oxygen Adsorption on Defective Graphene-Supported Pt Nanoparticles. *J. Phys. Chem. C.* 115, 22742-22747 (2011).

38. N. T. Kho, S. W. Kim, D. –H. Yoo, E. J. Kim, S. H. Hahn, Size-dependent work function and catalytic performance of gold nanoparticles decorated graphene oxide sheets. *Applied Catalysis A: General.* 469, 159-164 (2014).

39. M. Censabella, V. Torrisi, G. Compagnini, M. G. Grimaldi, F. Ruffino, Fabrication of metal nanoparticles-graphene Nanocomposites and study of the charge transfer effect. *Physica E:Low-dimensional systems and Nanostructures*. 118, 113887 (2020).

40. M. L. D. O. Pereira, R. D. S.Paiva, T. L. Vasconcelos, A. G. Oliveira, S. M. Oliveira, H. E. Toma, D. Grasseschi, Photoinduced electron transfer dynamics of AuNPs and Au@pdNPs supported on graphene oxide probed by dark-field hyperspectral microscopy. *Dalton trans.* 49, 16296 – 16304 (2020).

41. N. Torabi, S. Rousseva, Q. Chen, A. Ashrafi, A. Kermanpur, R. C. Chiechi, Graphene oxide decorated with gold enables efficient biophotovoltaic cells incorporating photo system I. *RSC Advances*. 12, 8783-8791 (2022).

42. D. –H. Lim, J. Wilcox, Mechanisms of the Oxygen Reduction Reaction on Defective Graphene Supported Pt Nanoparticles from First-Principles. *J. Phys. Chem. C*. 116, 3653-3660 (2012).

43. X. Fu, F. Bei, X. Wang, S. O'Brien, J. R. Lombardi, Excitation profile of surface-enhanced Raman Scattering in graphene-metal nanoparticle based derivatives. *Nanoscale*. 2, 1461-1466 (2010).

44. A. C. T. Van Duin, S. Dasgupta, F. Lorant, W. A. Goddard, ReaxFF: A Reactive Force Field for Hydrocarbons. *J. Phys. Chem. A*. 105, 41, 9396-9409 (2001).

45. S.Plimpton, Fast Parallel Algorithms for Short-Range Molecular Dynamics. *J Comp Phy*. 117, 1-9 (1995).

46. J. –W. Park, J. S. Shumaker-Parry, Structural Study of Citrate Layers on Gold Nanoparticles: Role of Intermolecular Interactions in Stabilizing Nanoparticles. *J. Am. Chem. Soc.* 136, 1907–1921 (2014).

47. S. G. Jiji, K. G. Gopchandran, Shape dependent catalytic activity of unsupported gold nanostructures for the fast reduction of 4-nitroanline. *Colloid and interface Science Communications.* 29, 9 -16 (2019).

48. V. Juvé, M. F. Cardinal, A. Lombardi, A. Crut, P. Maioli, J.Pérez-Juste, L. M. Liz-Marzán, N. Del Fatti, F. Vallée, Size-dependent surface plasmon resonance broadening in nonspherical nanoparticles: Single gold nanorods. *Nano Lett.* 13, 2234 – 2240 (2013).

49. L. G. Verga, J. Aarons, M. Sarwar, D. Thompsett, A. E. Russell, C. –K. Skylaris, Effect of graphene support on large Pt nanoparticles. *Phy. Chem. Chem. Phy.* 18, 32713-32722 (2016).

50. S. Karimi, A. Moshaii, S. Abbasian, M, Nikkhah, Surface plasmon resonance in Small gold nanoparticles: Introducing a Size-Dependent Plasma Frequency for Nanoparticles in quantum regime. *Plasmonics.* 14, 851-860 (2019).

51. X. Zhang, W. Chen, G. Wang, Y. Yu, S. Qin, J. Fang, F. Wang, X. –A. Zhang, The Raman red-shift of graphene impacted by gold nanoparticles. *AIP adv.* 5, 057133 (2015).

52. M. A. Bratescu, N. Saito, Charge doping of Large-Area Graphene by Gold-Alloy Nano particles. *J.Phys. Chem. C.* 117, 26804-26810 (2013).

53. G. Giovannetti, P. A. Khomyakov, G. Brocks, V. M. Karpan, J. van den Brink, P. J. Kelly, Doping graphene with metal contacts. *Phys. Rev. Lett.* 101, 026803 (2008).

54. A. K. Rappe, W. A. Goddard, Charge equilibration for molecular dynamics simulations. *J. Phys. Chem.* 95, 3358-3363 (1991).

# Supporting Information

# For

**Gold Nanoparticles Aggregation on graphene Using Reactive Force Field: A Molecular Dynamic Study**

J. Hingies Monisha[1, a], V. Vasumathi[1, b] and Prabal K Maiti[2]

[1]PG & Research Department of Physics, Holy Cross College, Tiruchirappalii-620002, Tamilnadu, India.

2 Center for Condensed Matter Theory, Department of Physics, Indian Institute of Science, Bangalore-560012, India.

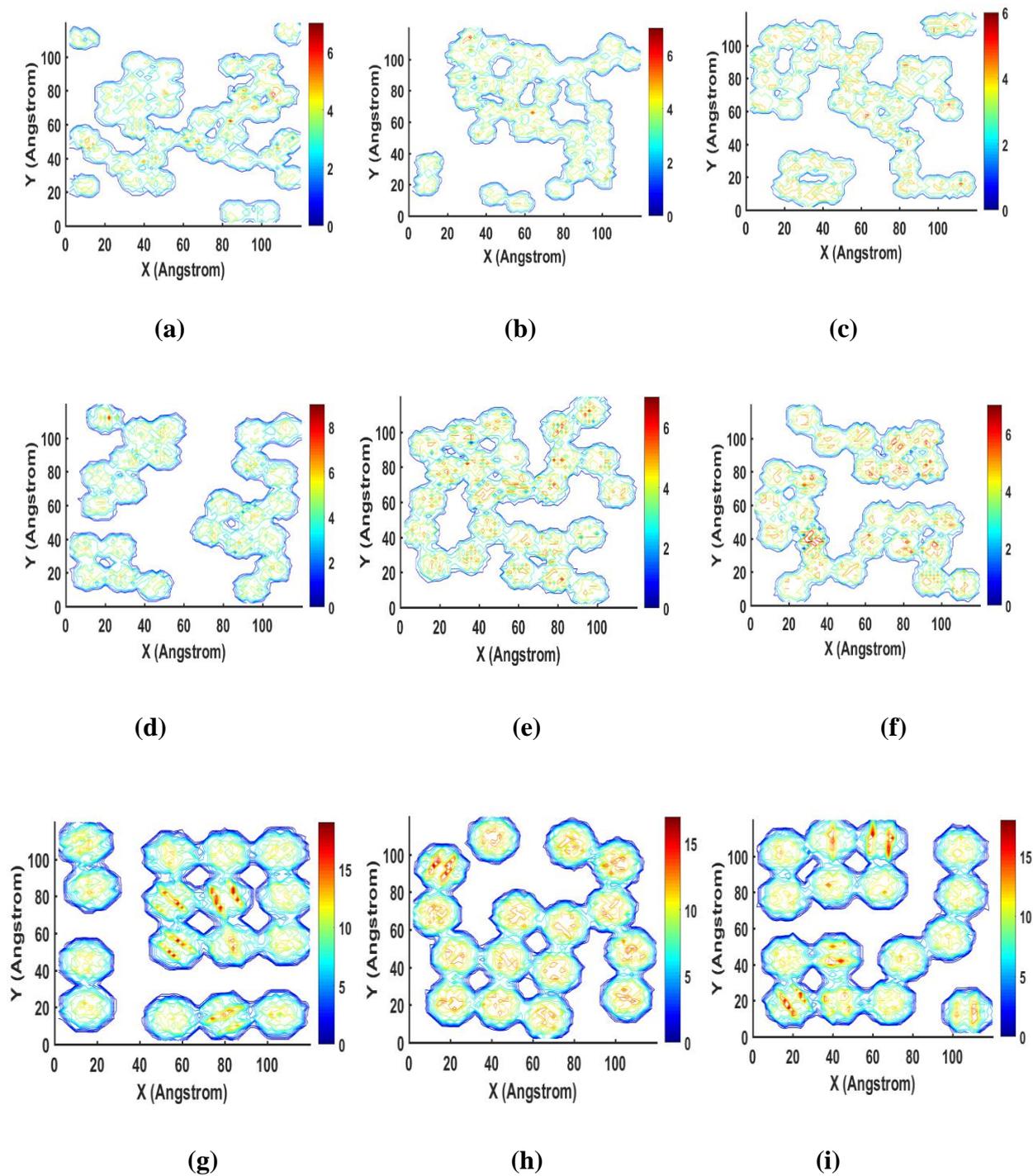

**Fig S. 1.** The contour map of GpAu-1.2 at (a) 300K, (b) 400K, (c) 500K, GpAu-1.6 at (d) 300K, (e) 400K , (f) 500K, GpAu-2.8 at (g) 300K, (h) 400K and (j) 500K.

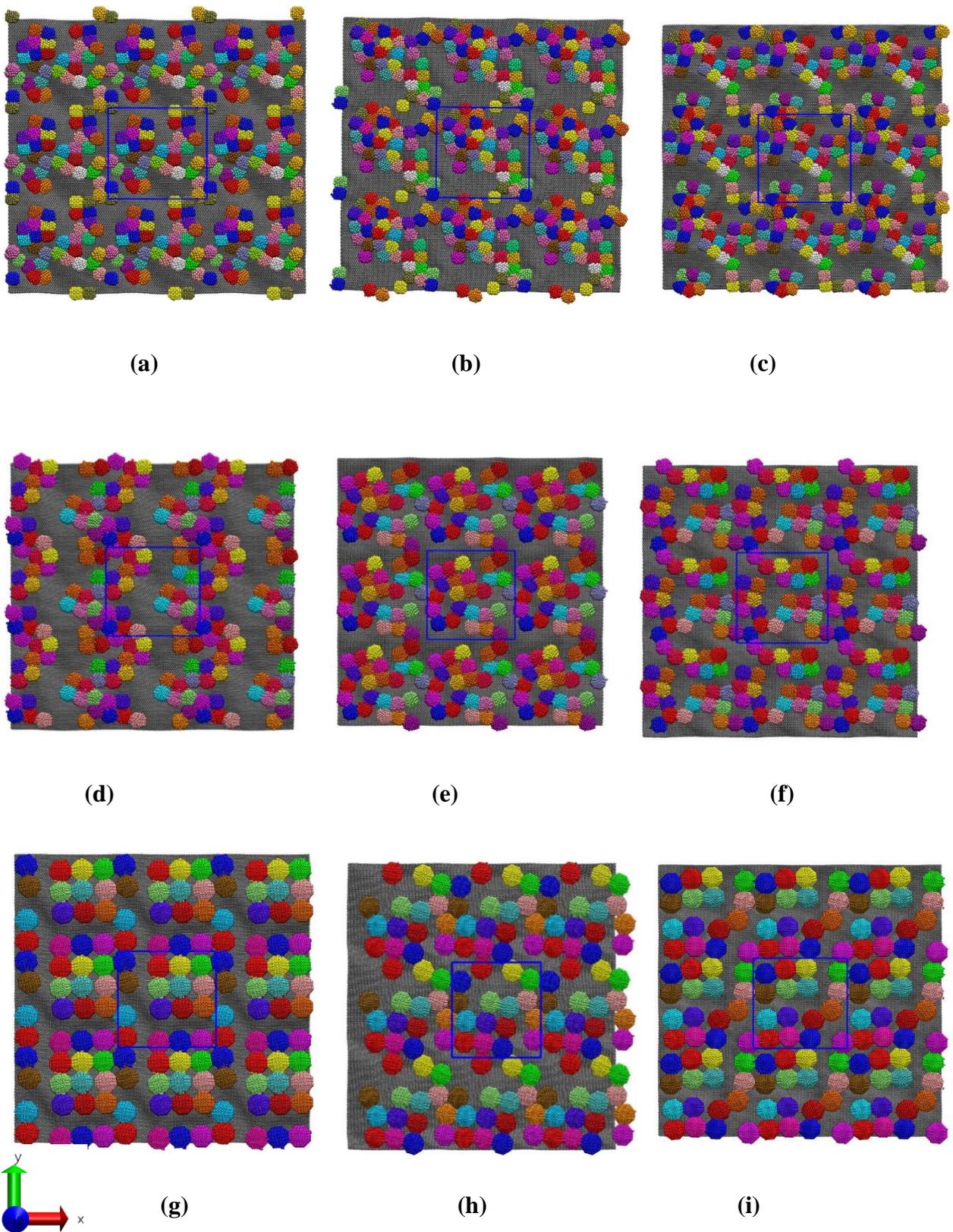

**Fig S. 2.** Snap shots of periodic view final configuration of GpAu-1.2 at (a) 300K, (b) 400K, (c) 500K, GpAu-1.6 at (d) 300K, (e) 400K , (f) 500K, GpAu-2.8 at (g) 300K, (h) 400K and (j) 500K.

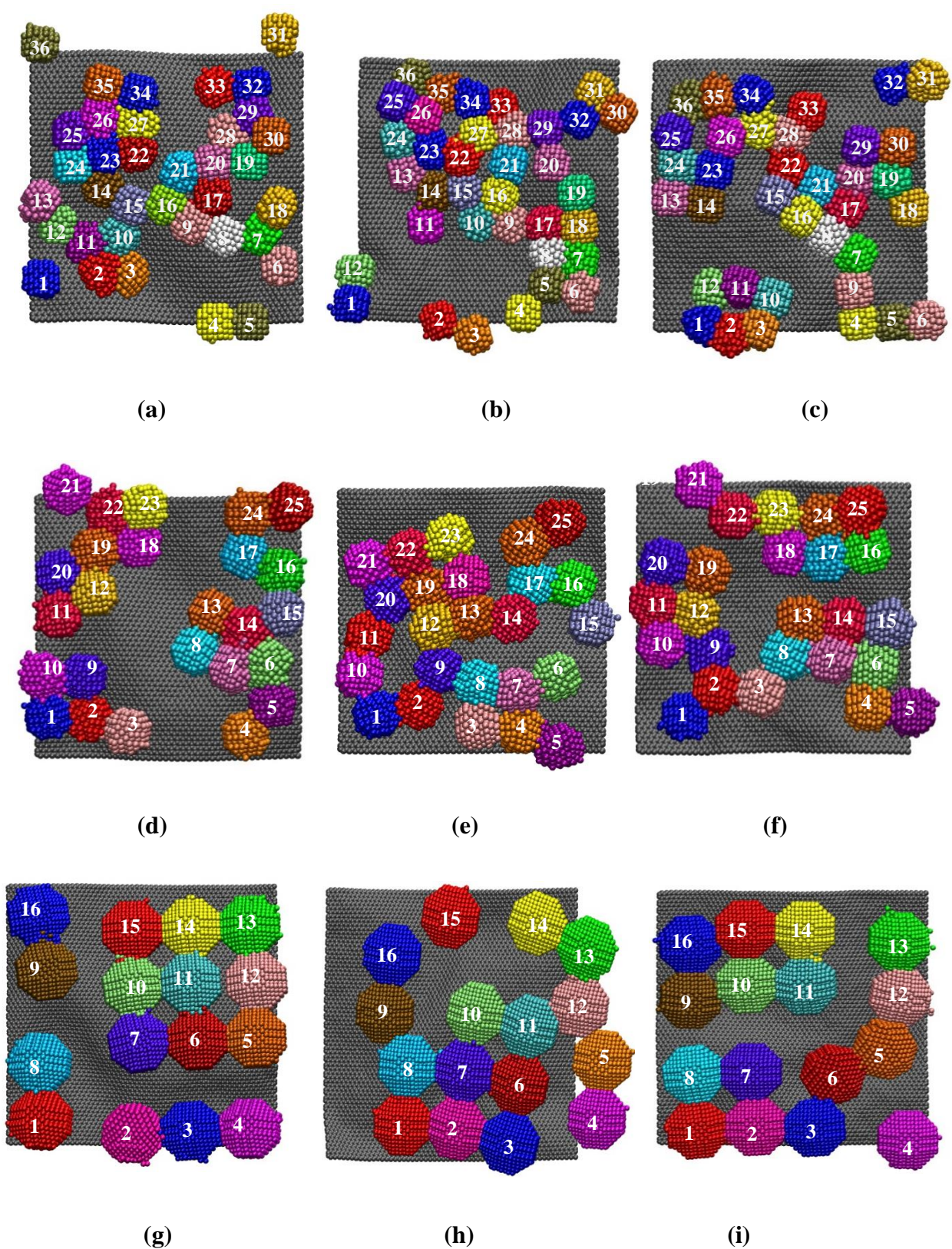

**Fig S. 3.** Snap shots of final configuration with ID number of AuNPs for GpAu-1.2 at (a) 300K, (b) 400K, (c) 500K, GpAu-1.6 at (d) 300K, (e) 400K , (f) 500K, GpAu-2.8 at (g) 300K, (h) 400K and (j) 500K.

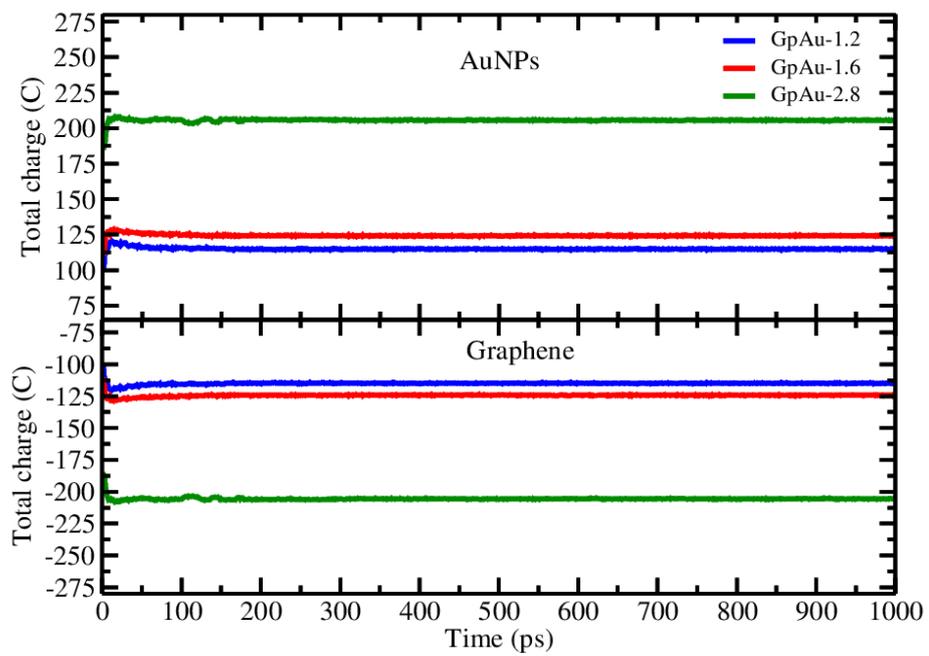

**Fig. S. 4.** Time evolution of total charge of AuNPs and graphene for all the systems at 300K.

**(a)**

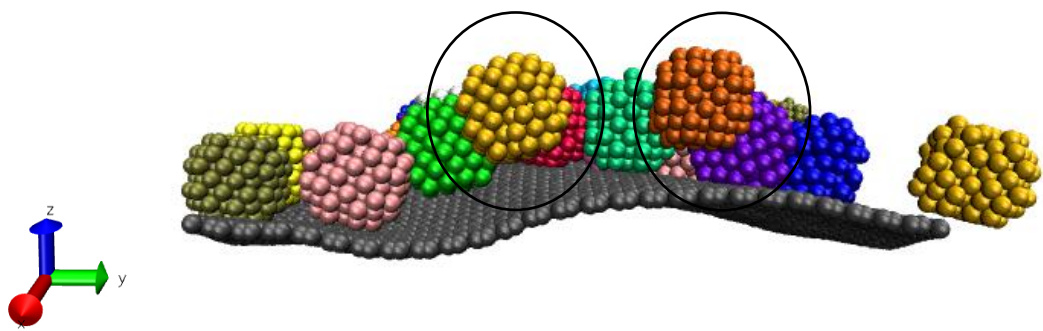

**(b)**

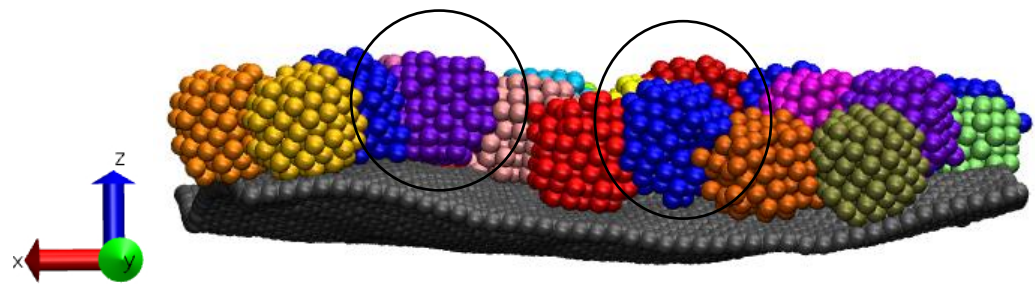

**(c)**

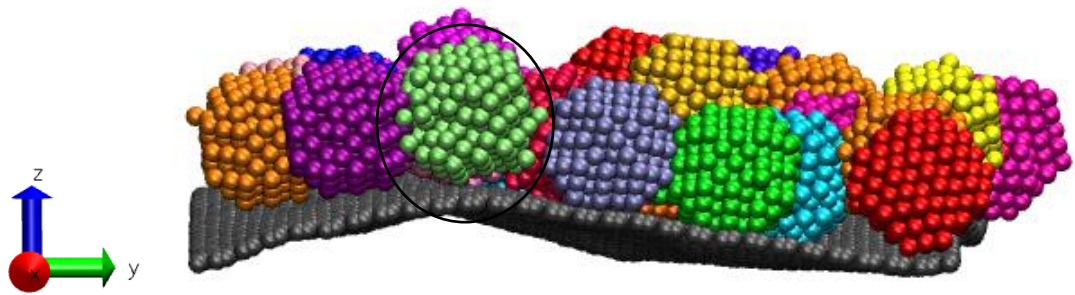

**(d)**

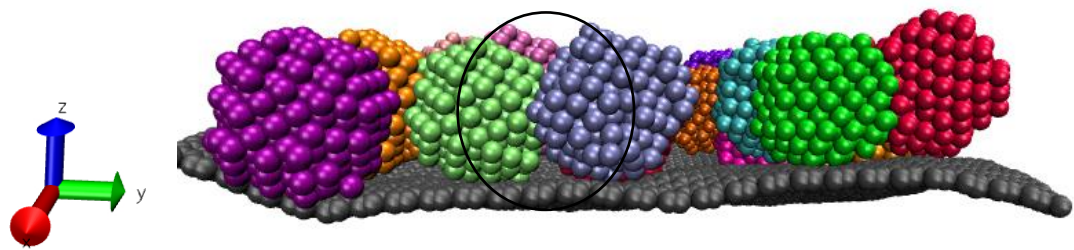

**(e)**

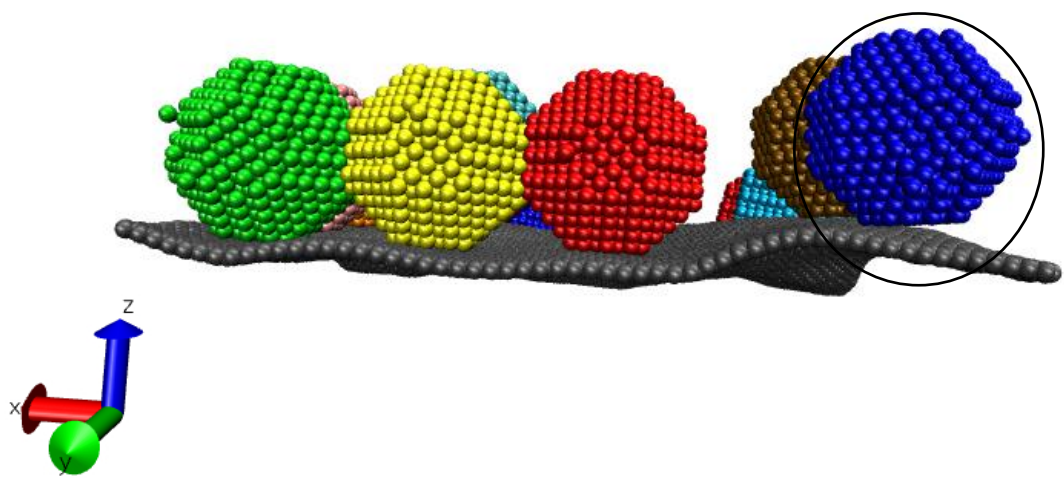

**Fig S. 5.** The snapshots of completely/partially elevated AuNPs from graphene for the system of GpAu-1.2 at (a) 300K, (b) 400K, GpAu-1.6 at (c) 300K , (d) 500K and GpAu-2.8 at (e) 300K.